\documentstyle[sprocl]{article}

\def\Journal#1#2#3#4{{#1} {\bf #2}, #3 (#4)}

\def\NPB{{\em Nucl. Phys.} B}
\def\PLB{{\em Phys. Lett.}  B}

\def\sc{\stackrel}

\def\sst{\scriptscriptstyle}

\def\be{\begin{equation}}
\def\ee{\end{equation}}
\def\bea{\begin{eqnarray}}
\def\eea{\end{eqnarray}}
\newcommand{\rpar}{\stackrel{\leftarrow}{\partial}}
\newcommand{\lpar}{\stackrel{\rightarrow}{\partial}}
\newcommand{\nn}{\nonumber\\}

\begin{document}
\renewcommand{\thefootnote}{\fnsymbol{footnote}}

\newpage
\setcounter{page}{0}
\pagestyle{empty}
\begin{flushright}
{August 2001}\\
{SISSA/64/2001/EP}\\
\end{flushright}
\vspace{3cm}

\title{EXTERIOR DIFFERENTIALS IN SUPERSPACE AND POISSON BRACKETS OF DIVERSE
GRASSMANN PARITIES\footnote{The talk at the 9th International Conference
on Supersymmetry and Unification of Fundamental Interactions (SUSY01),
(11-17 June 2001, Dubna, Russia)}}

\author{D. V. SOROKA}

\address{Kharkov Institute of Physics and Technology, 1 Akademicheskaya
Street,\\ 61108 Kharkov, Ukraine}

\author{V. A. SOROKA\footnote{On leave of absence from Kharkov Institute
of Physics and Technology, 61108 Kharkov, Ukraine}}

\address{International School for Advanced Studies (SISSA/ISAS)\\
Via Beirut 2--4, 34014 Trieste, Italy\\E-mail: vsoroka@kipt.kharkov.ua}


\maketitle\abstracts{It is shown that two definitions for the exterior
differential in superspace, giving the same exterior calculus, when
applied to the Poisson bracket lead to the different results. Examples
of the even and odd linear brackets, corresponding to semi-simple Lie
groups, are given and their natural connection with BRST and anti-BRST
charges is indicated.}

\newpage
\pagestyle{plain}
\renewcommand{\thefootnote}{\arabic{footnote}}
\setcounter{footnote}{0}

{\bf 1.\/} There exist two possibilities to define an exterior
differential in superspace with coordinates $z^a$ having Grassmann
parities $g(z^a)\equiv g_a$ and satisfying the permutation relations
\bea
z^az^b=(-1)^{g_ag_b}z^bz^a.
\eea
The first one realized when we set the Grassmann parity of the exterior
differential $d_{\sst0}$ to be equal to zero $g(d_{\sst0}z^a)=g_a$
and a symmetry property of an exterior product of two differentials is
\bea
d_{\sst 0}z^a\wedge d_{\sst 0}z^b=
(-1)^{g_ag_b+1}d_{\sst0}z^b\wedge d_{\sst0}z^a.
\eea
Another possibility arises when the parity of the differential $d_{\sst1}$
is chosen to be equal to unit $g(d_{\sst1}z^a)=g_a+1$ and the symmetry
property of the exterior product for two differentials is defined as
\bea
d_{\sst1}z^a\wedge d_{\sst1}z^b=
(-1)^{(g_a+1)(g_b+1)}d_{\sst1}z^b\wedge d_{\sst1}z^a.
\eea
In this case the symmetry properties of the exterior and usual products
of two differentials coincide
\bea
d_{\sst1}z^ad_{\sst1}z^b=
(-1)^{(g_a+1)(g_b+1)}d_{\sst1}z^bd_{\sst1}z^a.
\eea
The equivalence of the exterior calculi obtained with the use of the above
mentioned different definitions for the exterior differential can be
established as a result of the direct verification by taking into account
relations (2) and (3).

{\bf2.\/} However, we show that application of these two differentials
leads to the different results under construction from a given Poisson
bracket with a Grassmann parity $\epsilon=0,1$
\bea\label{in}
\{A,B\}_\epsilon=A\rpar_{z^a}{\sc{\epsilon}{\omega}}^{ab}(z)\lpar_{z^b}B
\eea
of another Poisson bracket (\ref{fin}).

Indeed, the corresponding to a Hamiltonian $H_\epsilon$
($g(H_\epsilon)=\epsilon$) Hamilton equations for the phase variables
$z^a$
\bea
{dz^a\over dt}=\{z^a,H_\epsilon\}_\epsilon=
{\sc{\epsilon}{\omega}}^{ab}\partial_{z^b}H_\epsilon
\eea
and the equations for their differentials
$d_{\sst\zeta}z^a\equiv y_{\sst\zeta}^a$ $(\zeta=0,1)$, obtained by
differentiation of this Hamilton equations, can be reproduced by the
following bracket of the Grassmann parity $\epsilon+\zeta$
\bea\label{fin}
(F,G)_{\epsilon+\zeta}=
F(\rpar_{z^a}{\sc{\epsilon}{\omega}}^{ab}\lpar_{y_{\sst\zeta}^b}&+&
(-1)^{\zeta(g_a+\epsilon)}\rpar_{y_{\sst\zeta}^a}
{\sc{\epsilon}{\omega}}^{ab}\lpar_{z^b}\nn&&{}+
\rpar_{y_{\sst\zeta}^a}y_{\sst\zeta}^c\partial_{z^c}
{\sc{\epsilon}{\omega}}^{ab}\lpar_{y_{\sst\zeta}^b})G
\eea
with the help of the Hamiltonian
$y_{\sst\zeta}^a\partial_{z^a}H_\epsilon$.
In the case $\zeta=1$, due to relations (3), (4), the terms in the
decomposition of a function $F(z^a,y_{\sst1}^a)$ into degrees $p$ of the
variables $y_{\sst1}^a$ can be treated as $p$-forms and the bracket
(\ref{fin}) can be considered as a Poisson bracket on $p$-forms such that
being taken between a $p$-form and a $q$-form results in a
$(p+q-1)$-form. The bracket (\ref{fin}) is a generalization of the bracket
introduced in~\cite{kar} on the superspace case and on the case of the
brackets (\ref{in}) with arbitrary Grassmann parities.

{\bf 3.\/} Now we apply this procedure to the linear even and odd brackets
connected with a semi-simple Lie group having structure constants
${c_{\alpha\beta}}^\gamma$. By taking as an initial bracket (\ref{in})
the linear even bracket given in terms of the commuting variables
$x_\alpha$ (here $z^a=x_\alpha$)
\bea
\{A,B\}_0=A\rpar_{x_\alpha}{c_{\alpha\beta}}^\gamma
x_\gamma\lpar_{x_\beta}B
\eea
and using the odd exterior differential $d_{\sst1}$, we obtain in
conformity with the transition from (\ref{in}) to (\ref{fin}) the
following odd linear bracket
\bea
(F,G)_1=F(\rpar_{x_\alpha}{c_{\alpha\beta}}^\gamma
x_\gamma\lpar_{\theta_\beta}+
\rpar_{\theta_\alpha}{c_{\alpha\beta}}^\gamma
x_\gamma\lpar_{x_\beta}+
\rpar_{\theta_\alpha}{c_{\alpha\beta}}^\gamma
\theta_\gamma\lpar_{\theta_\beta})G,
\eea
where $\theta_\alpha=d_{\sst1}x_\alpha$ are Grassmann variables. This
bracket has a nilpotent Batalin-Vilkovisky $\Delta$-operator~\cite{bv}
\bea
\Delta=-{1\over2}[\partial_{x_\alpha}(x_\alpha,\ldots)_1-
\partial_{\theta_\alpha}(\theta_\alpha,\ldots)_1]=
(T_\alpha+{1\over2}S_\alpha)\partial_{\theta_\alpha},\quad
\Delta^2=0,
\eea
where
\bea
T_\alpha={c_{\alpha\beta}}^\gamma x_\gamma\partial_{x_\beta},\quad
S_\alpha={c_{\alpha\beta}}^\gamma\theta_\gamma\partial_{\theta_\beta},
\quad Z_\alpha=T_\alpha+S_\alpha=(\theta_\alpha,\ldots)_1
\eea
are generators of the Lie group in the co-adjoint representation which
obey the commutation relations
\bea
[T_\alpha,T_\beta]={c_{\alpha\beta}}^\gamma T_\gamma,\quad
[S_\alpha,S_\beta]={c_{\alpha\beta}}^\gamma S_\gamma,\quad
[T_\alpha,S_\beta]=0.
\eea

By taking as an initial bracket (\ref{in}) the linear odd bracket given in
terms of Grassmann variables $\theta_\alpha$~\cite{s} (in this case
$z^a=\theta_\alpha$)
\bea
\{A,B\}_1=A\rpar_{\theta_\alpha}{c_{\alpha\beta}}^\gamma
\theta_\gamma\lpar_{\theta_\beta}B,
\eea
with the help of the differential $d_{\sst1}$ we come to the even linear
bracket of the form
\bea
(F,G)_0=F(\rpar_{\theta_\alpha}{c_{\alpha\beta}}^\gamma
\theta_\gamma\lpar_{x_\beta}&+&
\rpar_{x_\alpha}{c_{\alpha\beta}}^\gamma
\theta_\gamma\lpar_{\theta_\beta}\nn&+&
\rpar_{x_\alpha}{c_{\alpha\beta}}^\gamma
x_\gamma\lpar_{x_\beta})G,
\eea
where $x_\alpha=d_{\sst1}\theta_\alpha$ are commuting variables. This
bracket instead of the nilpotent second order differential
$\Delta$-operator has a nilpotent differential operator of the first order
\bea
Q={1\over2}[\theta^\alpha(x_\alpha,\ldots)_0-
x^\alpha(\theta_\alpha,\ldots)_0]=
\theta^\alpha(T_\alpha+{1\over2}S_\alpha),\quad Q^2=0.
\eea
If we consider $\theta^\alpha$ and $\partial_{\theta^\alpha}$ as
representations for the ghosts and antighosts respectively, then $Q$ and
$\Delta$ can be treated as the BRST and anti-BRST charges correspondingly
(see, e.g.,~\cite{h}) and satisfy the anticommutation relation
\bea
\{Q,\Delta\}={1\over2}(T^\alpha T_\alpha+Z^\alpha Z_\alpha),
\eea
two terms in the right-hand side of which, because of the commutation
relations
\bea
[T^\alpha T_\alpha,Q]=0, [T^\alpha T_\alpha,\Delta]=0,
[Z_\alpha,Q]=0, [Z_\alpha,\Delta]=0, [T^\alpha T_\alpha,Z_\beta]=0,
\eea
are central elements of the Lie superalgebra formed by the quantities $Q$,
$\Delta$, $T^\alpha T_\alpha$ and $Z^\alpha Z_\alpha$. $Z^\alpha Z_\alpha$
contains the term
\bea
D=\theta^\alpha\partial_{\theta^\alpha},
\eea
that can be considered as a ghost number operator which has the following
relations with $Q$ and $\Delta$:
\bea
[D,Q]=Q,\quad[D,\Delta]=-\Delta
\eea
and commutes with the central elements $T^\alpha T_\alpha$ and
$Z^\alpha Z_\alpha$
\bea
[D,T_\alpha]=0,\quad[D,Z_\alpha]=0.
\eea
Note that the Lie superalgebra for the quantities $Q$, $\Delta$, $D$,
$T^\alpha T_\alpha$ and $Z^\alpha Z_\alpha$ determined by the relations
(10), (15)--(20) can be used for the calculation of the BRST operator
cohomologies~\cite{hv}.

\section*{Acknowledgments}
One of the authors (V.A.S.) is grateful to  A.P. Isaev, J. Lukierski,
M. Tonin and J. Wess for useful discussions and to S.J. Gates Jr., P. Van
Nieuwenhuizen, W. Siegel and B. Zumino for stimulating discussions and
for hospitality respectively at the University of Maryland, SUNY (Stony
Brook) and LBL (Berkeley) where the parts of the work have been performed.
V.A.S. thanks L. Bonora for fruitful discussions and for hospitality at
SISSA (Trieste) where this work has been completed.

\end{document}